\documentstyle[epsbox]{ptptex}
\def\nue{\nu_{e}}
\def\num{\nu_{\mu}}

\def\nmnt{$\nu_{\mu}\leftrightarrow\nu_{\tau}$~}
\def\nenm{$\nu_{e}\leftrightarrow\nu_{\mu}$~}
\def\lsim{\lower.7ex\hbox{${\buildrel < \over \sim}$}}
\def\gsim{\lower.7ex\hbox{${\buildrel > \over \sim}$}}
\nofigureboxrule
\notypesetlogo
\title{K2K (KEK to Kamioka) neutrino-oscillation experiment
at KEK-PS\footnote{talk at the YITP workshop on flavor physics,
Kyoto, January 28-30, 1998}
}

\author{%
Yuichi Oyama
\footnote{E-mail address: oyamay@kekvax.kek.jp}\\
for K2K collaboration
\footnote{K2K collaboration includes physicists from KEK,
ICRR, Kobe, Niigata, Okayama, Tohoku, Tokai, Boston,
U.C.Irvine, Hawaii, LANL, SUNY, Washington, Chonnam,
Dongshin, Korea and SNU}\\
}

\inst{
Institute of Particles and Nuclear Studies,\\
High Energy Accelerator Research Organization,\\
Tsukuba, Ibaraki 305-0801 Japan
}

\abst{
A long-baseline neutrino-oscillation experiment using
a well-defined neutrino beam is in preparation at KEK.
Neutrinos generated at KEK will be
detected by the Super-Kamiokande detector
250~km away.
The design of the neutrino beam line,
beam monitors and detector are briefly presented.
The sensitivities for neutrino oscillations
are also discussed. 
}

\begin{document}

\maketitle

\section{Overview}
In recent years, several underground neutrino
observatories\cite{Kamioka,IMB,Soudan}
have reported that the atmospheric neutrino ratio,
$R \equiv (\num + {\bar \num})/(\nue + {\bar \nue})$,
is significantly smaller than the theoretical expectations.
This \char'134 atmospheric neutrino anomaly'' can be
explained by a neutrino oscillation hypothesis with an
oscillation parameter region of
$\Delta m^{2} \sim 10^{-2}$eV$^{2}$
and $\sin^{2} 2\theta \sim 1$.
The atmospheric neutrino anomaly has also been confirmed by
a preliminary result from the Super-Kamiokande
experiment,\cite{SuperK},
which shows a rather smaller $\Delta m^{2}$ than
does Kamiokande.

The K2K experiment\cite{Nishikawa}
(formerly called the KEK-PS E362 experiment)
is the first long-baseline neutrino-oscillation experiment
using an artificial neutrino beam. The almost pure
$\nu_{\mu}$ beam from $\pi^{+}$ decays is generated
in the KEK 12-GeV Proton Synchrotron, and is directed toward
the Super-Kamiokande detector, which is about 250km away
from KEK. The neutrino events observed in the
Super-Kamiokande detector are compared with neutrino events
in the front detector constructed at the KEK
site.
The nominal sensitive region on the neutrino-oscillation
parameter is $\Delta m^{2}$ =
$10^{-2}$eV$^{2} \sim 10^{-3}$eV$^{2}$, which covers the
parameter region suggested by the atmospheric neutrino
anomaly.
 
\section{Neutrino beam}

A neutrino beamline is under construction at
KEK.\cite{GPS} A picture of the neutrino beamline
under construction is shown in Figure~1.

\begin{figure}[t]
%\epsfysize=8. cm
%\centerline{\epsffile{beamline.eps}}
\centerline{\epsfile{file=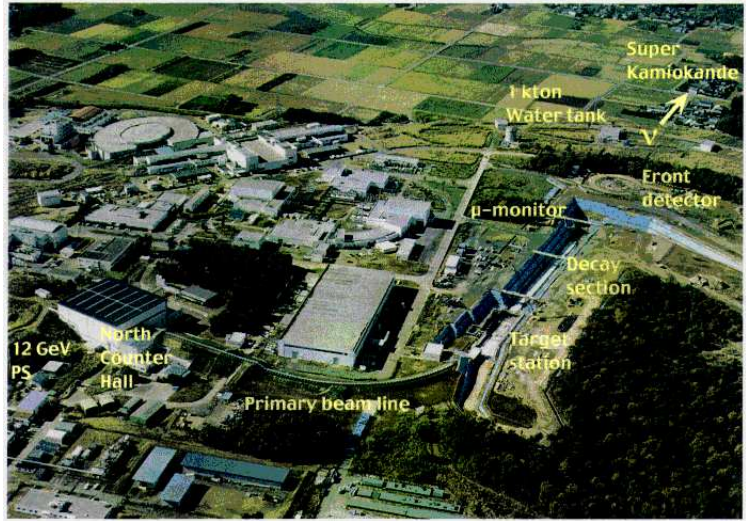,height=7.cm}}
%\centerline{\epsfig{beamline.eps,height=7.cm}}
\caption{Neutrino beamline under construction (November 19th, 1997).}
\end{figure}

A proton beam of 12~GeV is extracted from the Proton
Synchrotron with a fast extraction mode. The time width
of one spill is 1.1$\mu$sec. The frequency of the beam is
1 spill per 2.2 seconds and the nominal intensity is about
$6\times 10^{12}$ protons/spill.
A total of $1 \times 10^{20}$ p.o.t (protons on target)
will be extracted in 3 years of operation.

The proton beam is bent by 94$^{\circ}$ by dipole magnets,
and transported to an aluminum target of 2~cm$\phi
\times$~65~cm. Positively charged particles produced in the
target are focused by a toroidal magnetic field produced by a pair of
Horns\cite{Horn} toward the direction of the Super-Kamiokande detector.
The nominal pulse current of the Horn is 250kA.
The muon neutrino flux will be enhanced by a factor of
14 by installing the Horns. A test operation
of the Horns was successfully completed in fall, 1997.

In a 200~m decay tunnel followed with the Horn magnets,
one $\pi^{+}$ decays to one $\nu_{\mu}$ and one $\mu^{+}$.
Muons and the remaining pions are absorbed in a beam dump
downstream of the decay tunnel.

The generation of the neutrino beam is simulated using
a Monte-Carlo simulation. The energy spectrum and angular
dependence of the beam at the front detector
(300~m downstream from the target)
and at the Super-Kamiokande detector (250~km downstream)
are shown in Figure~2. 
The mean energy of the neutrino beam is about 1.4~GeV,
and the peak energy is about 1~GeV.
The contamination of electron neutrinos was calculated to be
$\sim 1\%$.
The energy spectrum of the neutrino beam is uniform
within an angular spread of 3~mrad from the center of
the beam axis.
Because the angular acceptance of the Super-Kamiokande detector
from the KEK site is $\sim$ 50m/250km = 0.2~mrad,
the divergence of the neutrino beam is much larger than
the size of the Super-Kamiokande detector.

\begin{figure}[t]
% \epsfysize=7.5 cm
% \centerline{\epsffile{hornflux.eps}}
\centerline{\epsfile{file=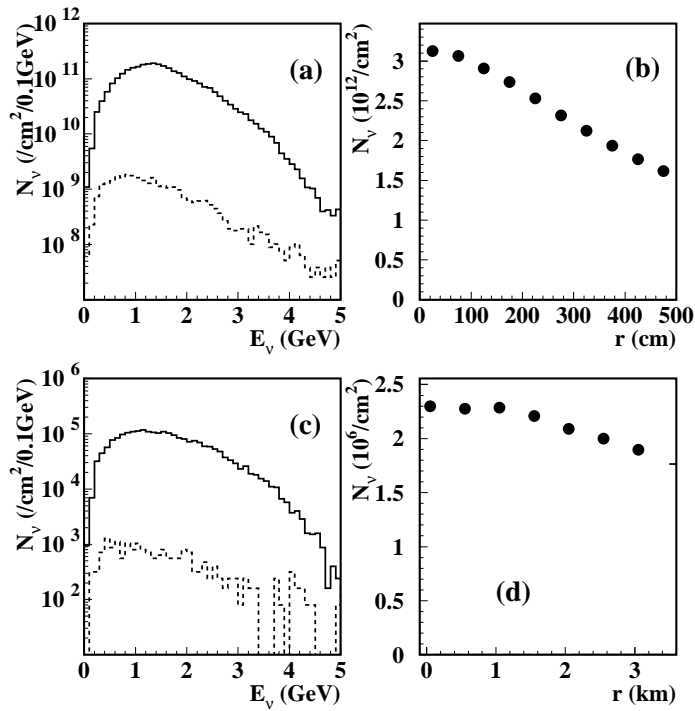,height=10.5cm}}
%\centerline{\epsfig{fluxplot.eps,height=10.5cm}}
\caption{Expected energy spectrum and radial distribution
for 10$^{20}$ protons on the target
(corresponding to 3 years of operation), both at the front detector
($L$ = 300~m) and Super-Kamiokande ($L$ = 250~km).}
\label{fig:hornflux}
\end{figure}

\section{Beam monitors}

In order to obtain the neutrino-energy spectrum,
a pion monitor\cite{Inagaki}
will be set downstream of the second Horn.
The pion monitor is a gas Cherenkov detector with a
spherical mirror and R-C318 gas.
Charged pions emit Cherenkov light in the gas, and
a Cherenkov ring is created in the focus plane.
The light intensity of the Cherenkov light is monitored
by ADC on a spill-by-spill basis.
The information concerning the Cherenkov-light intensity and shape
of the ring are used to calculate the momentum distribution and 
divergence of the pion beam. The ratio of neutrino flux
at the front detector and at Super-Kamiokande site can be
expected for a neutrino energy larger than 1.0~GeV
based on the decay kinematics of the pions.

A muon monitor will be installed downstream
of the decay tunnel.
The muon monitor is a 2~m $\times$ 2~m pad-type ionization  chamber
filled with He gas. The anode current induced by the muon beam is
read out from the x-direction and the y-direction with an interval
of 5~cm. The 2-dimensional projection of the muon intensity
provides information about the beam profile.
The position of the beam center is obtained with
an accuracy of $\sim$2~cm.  

\section{Front detector}

The front detector in the KEK site is located 300~m
downstream of the target. A schematic view of the front
detector is shown in Figure~3.
We will construct two types of detectors. They are a 1kt water
Cherenkov detector and a so-called Fine-Grained Detector.

\begin{figure}[t]
\centerline{\epsfile{file=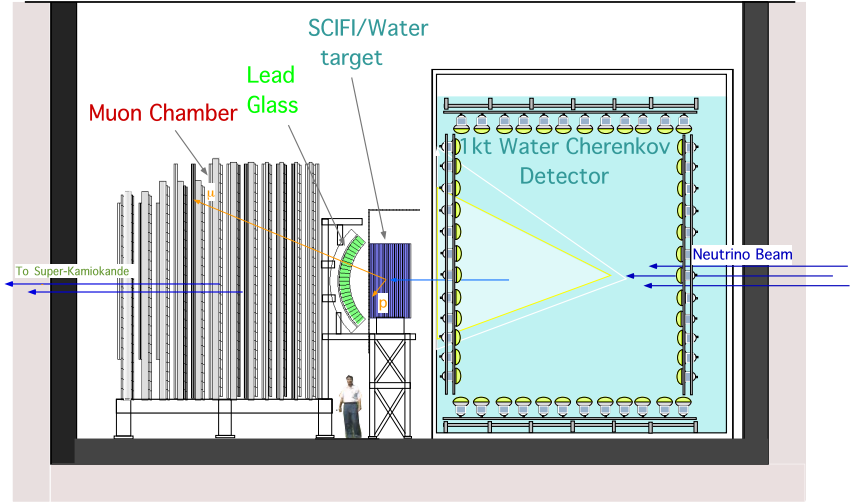,height=7.0cm}}
%\centerline{\epsfig{detect.eps,height=7.0cm}}
 \caption{Front detector in the K2K experiment.
          The cylindrical experimental 
          hall is 24m in diameter and 16m deep.}
\label{fig:neard}
\end{figure}

The 1kt water Cherenkov detector is a miniature of the
Super-Kamiokande detector. The detector volume of 496~tons
is viewed by 860 20-inch photomultiplier tubes with a 
70~cm spacing. The main purpose of the 1kt water Cherenkov
detector is to compare neutrino events in the KEK site with
events in the Super-Kamiokande detector using the same
observation method. A direct comparison of the neutrino
events cancels any systematic errors inherent to
water Cherenkov detectors.

The purpose of the Fine-Grained Detector is to
understand the neutrino-flux profile and energy distribution
precisely at the KEK site. The Fine-Grained Detector
consists of 4 detector
elements: a scintillating fiber tracker, trigger
counters, lead glass counters and a muon ranger.

\begin{figure}[t]
\centerline{\epsfile{file=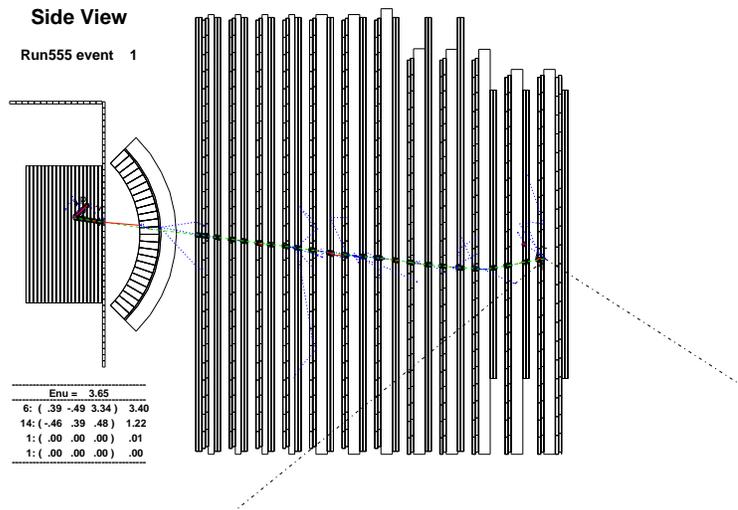,height=8.0cm}}
%\centerline{\epsfig{mus.eps,height=8.0cm}}
 \caption{Cross-sectional view of a typical
          quasi-elastic scattering
          event in the Fine-Grained Detector
          obtained from a Monte-Carlo simulation}
\label{fig:event}
\end{figure}

The scintillating fiber tracker is a 20-layer
\char'134 sandwich" of scintillating fiber sheets and
a water target. One scintillating fiber layer consists of 4
scintillating fiber sheets (2 vertical and 2 horizontal)
with a sensitive area of 2.4~m $\times$ 2.4~m.
They are made of 0.7~mm$\phi$ scintillating fibers.
The scintillating fiber layers are arranged with intervals
of 9cm. The water target is contained in 1.8~mm-thick
aluminum tubes, and are placed between scintillation fiber layers.
The scintillation light from the fibers is read by
Image Intensity Tubes (IITs) and CCD cameras.
From a cosmic-ray muon test,
the detection efficiency of each layer was found
to be better than 99\%, and the position resolution of the
fiber sheet was obtained to be $\sim$~280~$\mu$m.
The data from the scintillating fiber tracker are used
to reconstruct the tracks of charged particles generated in
neutrino interactions and to identify the kinematics of the
neutrino interaction.

Around the scintillating fiber tracker, 
about 100 large plastic scintillators ($\sim$~4~m (length)
$\times \sim$10~cm (width) $\times$ 4~cm (thickness)) from
the previous Trinstan experiments are aligned. The absolute
time of an event is obtained by plastic scintillators
with an accuracy of 0.7~nsec.
This information is used to reject cosmic-ray muon
background with a reduction rate of $\sim$~99\%, and is
used in the time correction of the muon chamber data.
Neutrino interactions outside of the Fine-Grained
Detector can also be identified by the counter.

The purpose of the lead-glass counters is to precisely
measure the contamination of electron neutrinos in the muon
neutrino beam. It consists of 600 lead-glass counters;
each counter is 12~cm $\times$~12~cm in acceptance.
The basic idea of the 
electron identification is based on the response of the
lead-glass counters to muons and electrons.
The energy deposit of electron can be measured with a
resolution of $8\% / \sqrt{E_{e}}$, whereas the energy
deposit of muons which pass through the lead-glass
counters is less than $\sim$~1.0GeV.
Therefore, high-energy ($\gsim$ 1.2GeV) electrons can
obviously be distinguished from muons.
The contamination of electron neutrinos can be measured
to be (1.0 $\pm$ 0.3)\% when the $\nue$ contamination
is 1\%. Calibration of the lead-glass counters using
cosmic-ray muons was finished in Dec. 1997.

The muon ranger consists of $\sim$~900 modules of drift
tubes from the Venus muon chamber and 12 plates of an iron
filter (10$\sim$20~cm thick). Muons generated by
charged-current interactions in the water target
are reconstructed with a spatial resolution of 2.2~mm.
The efficiency of each drift tube layer is $\sim$~99\%.
The muon momentum is calculated from the total depth
of the material (mainly iron) between the position
of the neutrino interaction and the position where the
muon is stopped. 
The muon ranger is used to obtain the muon energy with a
resolution of $\Delta E_{e} / E_{e} = 8 \sim 10\%$.

A schematic view of a typical neutrino interaction in the
Fine Grained detector is shown in Figure~4.

A detailed description of the Super-Kamiokande,
which will be used as a far detector, has been
made,\cite{SuperK,Nishijima} and is not presented here.

The performance of the 1kt water Cherenkov detector,
Fine-Grained Detector and Super-Kamiokande are summarized in
Table.I. Number of neutrino events expected in the fiducial
volume are also shown in Table.II.

\begin{table}[t]
\vskip 1cm
\caption{Summary of the detector performance in the K2K experiment.
\label{tab:evrate}}
\begin{center}
\begin{tabular}{l|ccc}
\hline
\hline
          & Fine-grained & 1kton  & Super-\\ 
    ~     & detector & detector & Kamiokande\\ 
\hline
$\Delta E_{\mu} / E_{\mu}$  &   8 $\sim$ 10 \%   &     &   3 \%  \\
$\Delta \theta_{\mu}$   &   $\sim 1^{\circ}$ & $3^{\circ}$ & $3^{\circ}$ \\
$\Delta V_{fid} / V_{fid}$  &  $\sim$ 1 \%   & $\sim$ 10 \%  &   3 \%  \\
$\Delta E_{e} / E_{e}$  & 8\%/$\sqrt{E_{e}}$ & 3\%/$\sqrt{E_{e}}$ & 3\%/$\sqrt{E_{e}}$ \\
\hline
\hline
\end{tabular}
\end{center}
\end{table}

\begin{table}[t]
\vskip 1cm
\caption{Number of neutrino interactions in the fiducial volume.}
\begin{center}
\begin{tabular}{l|rrr}
\hline
\hline
          & Fine-grained & 1kton  & Super-\\ 
    ~     & detector & detector & Kamiokande\\ 
\hline
fiducial volume &   5.8 ton  & 21 ton    &  22500 ton  \\
~~~~(dimension)   &  (2m$\times$2m$\times$1.1m) & (3m$\phi \times$ 3m) & (34m$\phi \times$ 32m) \\
neutrino flux&$3.0\times 10^{12}$/cm$^{2}$ &$3.0\times 10^{12}$/cm$^{2}$&$2.3\times 10^{6}$/cm$^{2}$\\
event rate&0.010/spill & 0.033/spill & $\sim$1/day\\
$\nu_{\mu}$ event & 171600 & 552000 & 465 \\
~~~~(quasi-elastic) & 44200 & 142000 & 120 \\
$\nu_{e}$ event & 1250 & 4000 & 4 \\
\hline
\hline
\end{tabular}
\end{center}
\end{table}

\section{Sensitivity}

An examination of the \nenm oscillation in K2K
is an appearance search.
Any small contamination of $\nue$ in the $\num$ beam
can be confirmed by the lead-glass counter.
An excellent particle-identification capability
in the Super-Kamiokande detector was already examined.\cite{Kasuga}
Therefore, a possible excess of electron neutrino events in
the Super-Kamiokande detector is direct evidence of the
neutrino oscillation.
Although there is a possible background of $\pi^{0}$
from a neutral-current interaction of $\num$,
it can be easily recognized, because most of
the $\pi^{0}$ have an energy of less than 1.0~GeV.
If the oscillation parameters, $(\Delta m^{2},\sin^{2}2\theta)
= (1 \times 10^{-2},1)$, are assumed, the number of
electron events with $E_{e} > 1.5$GeV is expected to be $\sim$90,
whereas the background from $\pi^{0}$ and from $\nu_{e}$
contamination is only $\sim$4. 

\begin{figure}
\centerline{\epsfile{file=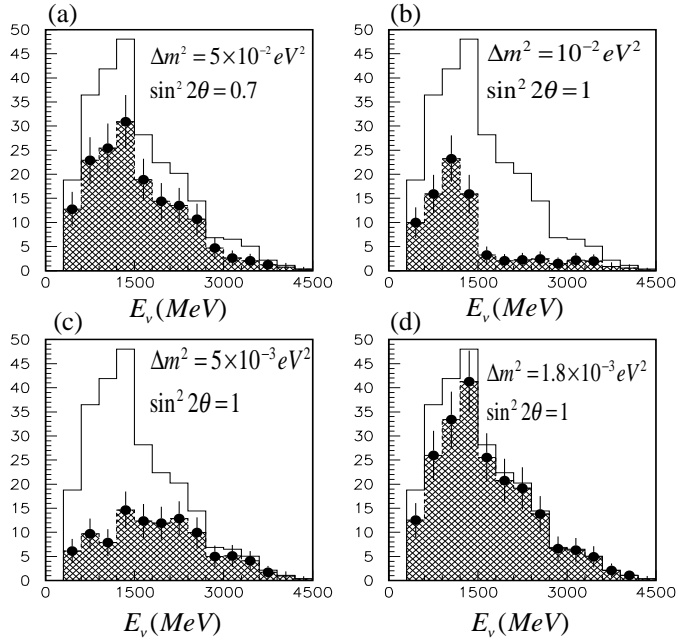,scale=0.47}}
%\centerline{\epsfig{spec.eps,scale=0.47}}
\caption{Expected $\nu_{\mu}$ energy spectra with
Super-Kamiokande at
$10^{20}$ protons on target for various \nmnt 
oscillation parameters (a)-(d) (data points with error bars) and for
the no oscillations (solid histogram). }
\label{f:nutau-test}
\end{figure}

The \nmnt oscillation can be examined by studying any
distortion of the neutrino energy spectrum at Super-Kamiokande.
The expected neutrino energy spectrum at given neutrino oscillation
parameters is shown in Figure~5. The change in the neutrino energy
spectrum as well as a reduction of neutrino events would comprise
an evidence of neutrino oscillation.

In the determination of the neutrino energy spectrum, we will employ
a quasi-elastic interaction of muon neutrinos,
because quasi-elastic scattering is recognized as single
ring events in the Super-Kamiokande detector, and because the energy
of the neutrinos can be calculated from the momentum and the scattering
angle of the muon track.
It should also be noted that the cross section of quasi-elastic
scattering is well understood compared with that of other interactions.

The sensitivity of the \nenm and \nmnt oscillation in K2K is shown
in Figure~6 together with other experiments
\cite{Kamioka,SuperK,CDHS,CHARM,MINOS,E776,Krasno,Bugey,LSND,KARMEN,Gosgen,Chooz}
which are proposed,
under construction, in data taking, or finished.
The sensitive parameter regions in K2K are
$\Delta m^{2} \gsim 1 \times 10^{-3}$eV$^{2}$ and $\sin^{2}2\theta > 0.1$
for the \nenm oscillation, and 
$\Delta m^{2} \gsim 3 \times 10^{-3}$eV$^{2}$ and $\sin^{2}2\theta > 0.4$
for the \nmnt oscillation.

\medskip

The first neutrino beam in K2K will be available in January, 1999. 

\begin{figure}[t]
\centerline{\epsfile{file=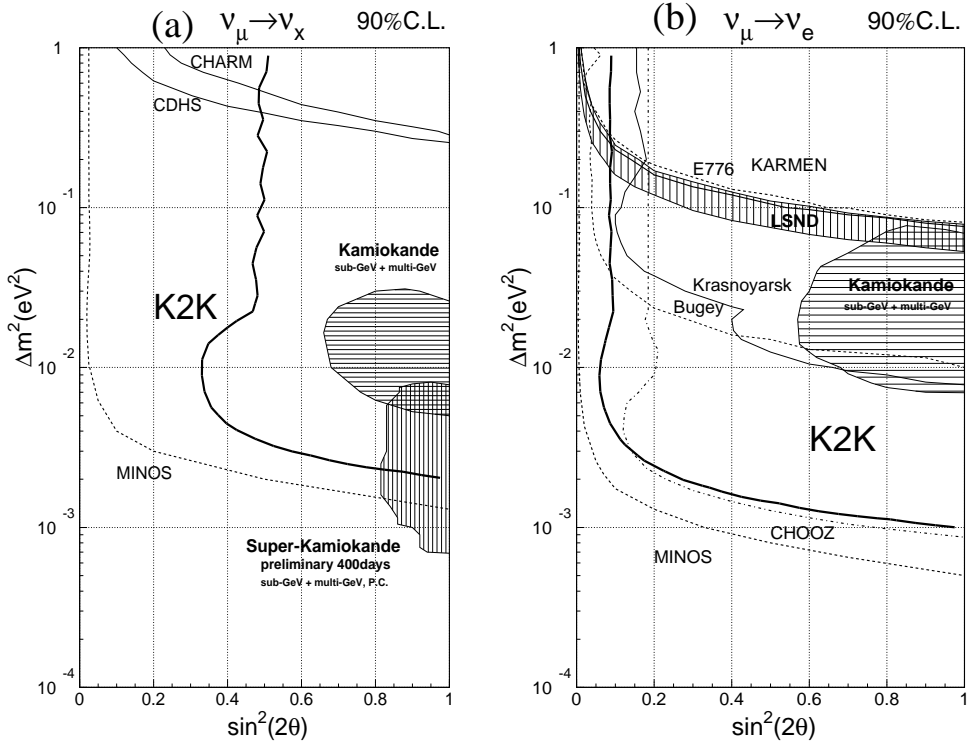,width=13.5cm}}
%\centerline{\epsfig{contour.eps,width=13.5cm}}
\vspace{10pt}
\caption{Exclusion contours of 90 \% C.L. for
(a)\nmnt oscillation and
(b)\nenm oscillation (thick solid lines).
The allowed regions by Kamiokande,\cite{Kamioka} LSND\cite{LSND}
and Super-Kamiokande\cite{SuperK} and excluded (or sensitive)
regions by other experiments are also
plotted.\cite{CDHS,CHARM,MINOS,E776,Krasno,Bugey,LSND,KARMEN,Gosgen,Chooz}.}
% SK has not yet
%presented analysis for $\nu_\mu\rightarrow\nu_e$.}
\label{fig:cont}
\end{figure}

\end{document}